\documentclass[aps,prb,twocolumn,groupedaddress,showpacs]{revtex4}

\usepackage{graphicx}

\begin{document}

\title{Energy Spectrum of Bloch Electrons
Under Checkerboard Field Modulations}

\author{Ming-Che Chang}
\affiliation{Department of Physics, National Taiwan Normal
University, Taipei, Taiwan}
\author{Min-Fong Yang}
\affiliation{Department of Physics, Tunghai University, Taichung,
Taiwan}

\date{\today}
\begin{abstract}
Two-dimensional Bloch electrons in a uniform magnetic field
exhibit complex energy spectrum. When static electric and magnetic
modulations with a checkerboard pattern are superimposed on the
uniform magnetic field, more structures and symmetries of the
spectra are found, due to the additional adjustable parameters
from the modulations. We give a comprehensive report on these new
symmetries. We have also found an electric-modulation induced
energy gap, whose magnitude is independent of the strength of
either the uniform or the modulated magnetic field. This study is
applicable to experimentally accessible systems and is related to
the investigations on frustrated antiferromagnetism.
\end{abstract}



\maketitle

\section{introduction}

When the spectrum of a two-dimensional (2D) Bloch electron in a
uniform magnetic field is plotted in the energy-flux diagram, a
self-similar structure with fractal property
emerges.\cite{hofstadter76} Such a complex structure, called the
Hofstadter spectrum, arises due to the commensurability between
two length scales in this system: the lattice constant and the
cyclotron radius. The Hofstadter spectrum is one of the earliest
predictions of fractal structure in solid-state physics.
Subsequently, it was found that not only the energy spectrum has
self-similarity, the wave function also exhibits scaling behavior
and can be analyzed using the renormalization
group.\cite{thouless83}

Because of its beautiful structure, the Hofstadter spectrum has
attracted many researchers' attention, and the spectra for
different 2D lattice symmetries have been reported. Besides the
square lattice, there are also triangular lattice,\cite{claro79}
honeycomb lattice,\cite{rammal85} Kagome lattice,\cite{xiao02} and
a bipartite periodic structure with hexagonal
symmetry.\cite{vidal98} These are all studied within the framework
of Bloch electrons in a uniform magnetic field, usually assuming
nearest-neighbor (NN) couplings $t_1$ only. Including and varying
the next-nearest-neighbor (NNN) couplings $t_2$ leads to
band-crossings accompanied by exchange of quantized Hall
conductances between bands.\cite{hatsugai90,lee94} For a square
lattice, detailed scaling analysis reveals a bicritical point at
$t_1=2t_2$, accompanied by interesting topological change of the
spectrum.\cite{han94} Spectra for systems with couplings beyond
next-nearest neighbors have also been studied.\cite{barelli92} In
addition, the external magnetic field, rather than being uniform,
can be periodically modulated with a pattern unrelated to the
original lattice. The simplest situation when a magnetic lattice
overlaps with the electric lattice is realized when a
ferromagnetic grid is deposited on a
semiconductor.\cite{gerhardts96} The interfacial stress between
two materials would naturally induce an electric grid with the
same period and symmetry as the ferromagnetic grid. More
generally, there can also be a magnetic modulation with the
pattern of a 1D strip,\cite{oh95} or a 2D
checkerboard\cite{shi97,oh99,ito99} superimposes on the electric
square lattice. The checkerboard configuration has been realized
experimentally using a superconducting Nb-network with periodic
magnetic Dy-islands.\cite{ito99} The calculations of the
Hofstadter spectra provide the basis to study such articifial
networks.

A direct observation of the Hofstadter spectrum has been realized
using microwaves\cite{kuhl98} or acoustic waves\cite{richoux02}
transmitting through an array of {\it macroscopic} scatters.
However, a fractal {\it electronic} spectrum is significantly more
difficult to be realized in an usual solid, whose lattice constant
is only a few angstroms, and a magnetic field of the order of
$10^4$ Teslas is required. For 10 Teslas or less, we can only
probe the part of the Hofstadter spectrum that reduces to the
familiar Landau levels with roughly equal spacings in energy. In
the last decade, different superlattice structures with much
larger lattice constants are used to cope with this high-field
problem.\cite{albrecht01} Besides, several physical systems are
closely related to the Hofstadter problems and offer alternative
angles of investigation, for example, the studies of a
superconductor in a vortex state,\cite{morita01} a superconducting
network in a magnetic field,\cite{ito99} and a junction of three
quantum wires.\cite{chamon03} Furthermore, recent advance on
optical lattices makes it possible to implement a lattice
Hamiltonian resembling the effects of magnetic fields with {\it
neutral} atoms.\cite{jaksch03} This offers great opportunities
since not only the magnetic field, but also the lattice symmetry,
the potential strength, and the relative importance of many-body
effects can be adjusted in such a system.

This paper is motivated by a study very different from those
mentioned above. In a recent paper,\cite{misguich01} Misguich and
coworkers, by using the hard-core bosons to represent the spin
degrees of freedom, and using the Chern-Simons transformation to
transmute bosons to fermions, mapped a 2D frustrated
antiferromagnetic problem to a Hofstadter problem. This approach
is subsequently used to study the magnetization properties of the
$J_1-J_2$ Heisenberg model on a square lattice.\cite{chang02}
After suitable mathematical mappings and a mean-field
approximation, the magnetization problem can be reduced to a
Hofstadter problem with {\it both} electric and magnetic
super-structures superimpose on the original lattice. This
motivates us to consider the checkerboard super-structure, which
is related to the N\'eel phase in the magnetization problem, with
congruous electric {\it and} magnetic modulations. Couplings up to
next nearest-neighbors are considered, which are essential to
cause magnetic frustration in the $J_1-J_2$ Heisenberg
model.\cite{note0}

In this paper, we make a comprehensive survey of the symmetries of
the Hofstadter spectra with field modulations. Some of the
symmetries already exist without modulations, such as the ones
related to reversing the direction of the magnetic flux, and
shifting the flux in a plaquette by two flux quanta (see items II
and III in Sec. II). Some of the other symmetries that are closely
related to field modulations are reported for the first time. In
particular, when the system is subject to staggered $\pi$-fluxes,
its spectrum in the $E-\phi$ diagram has an up-dowm symmetry even
with NNN couplings (see Fig.~5), which is quite unexpected since
NNN couplings usually destroy such a symmetry.\cite{hatsugai90}
Besides the studies on symmetries, for systems with only NN
couplings, we find a simple algebraic connection between the
spectra with and without the {\it electric} checkerboard field. We
also find a flux-independent energy gap induced by the electric
modulation (see Fig.~3), which can be explained by using the
algebraic relation just mentioned.

This paper is organized as follows. Theoretical formulation on the
system with checkerboard super-structure, as well as the
discussion of the symmetries of the energy spectra, can be found
in Sec. II. Major features of the Hofstadter spectra are discussed
in Sec. III. We summarize and conclude our results in Sec. IV. The
proofs on checkerboard-translation symmetry of the spectrum and on
the existence of the flux-independent energy gap are given in the
appendices.

\section{theoretical analysis}

The tight-binding Hamiltonian describing the motion of an electron
in a magnetic field is given by
\begin{equation}
H = -\frac{1}{2} \sum_{\left\langle i,j\right\rangle} \left(
t_{ij} e^{i\theta_{ij}} \, f^+_i
 f_j + h.c. \right) + \sum_i V_i f^+_i f_i \; , \label{tight-binding}
\end{equation}
where $\theta_{ij}$ is the magnetic phase factor. For
clarification, we will replace the label $i$ by $(n, m)$ in the
following, which denotes the $(n, m)$ plaquette as well as the
lattice point at the lower left corner of the plaquette. Without
loss of generality, we take the uniform part of the magnetic flux
through plaquette as $\phi=2\pi p/q$ with relative prime integers
$p$ and $q$. For the checkerboard modulation, we have
$\delta\phi_i / 2\pi = - \Delta_\phi (-1)^{n+m}$ and $V_i \equiv
\Delta_V (-1)^{n+m}$ (see Fig.~1). The Landau gauge is used such
that the magnetic phase factors become
\begin{equation}
\left\{
\begin{array}{l}
\theta_{n+1,\;m;\;n,\;m}=0 \; , \nonumber \\
\theta_{n,\;m+1;\;n,\;m}=n\phi +(-1)^{n+m} \pi \Delta_\phi \; , \\
\theta_{n+1,\;m+1;\;n,\;m}=\theta_{n,\;m+1;\;n+1,\;m}= \left( n +
{1 \over 2} \right) \phi \; . \nonumber
\end{array}\right.     \label{theta_2D}
\end{equation}

Due to the modulation in the $y$ direction and under the gauge
choice in Eq.~(\ref{theta_2D}), the tight-binding Hamiltonian in
Eq.~(\ref{tight-binding}) now becomes invariant under the
$y$-translation $m \rightarrow m+2$. Thus the Bloch theorem gives
\begin{equation}
f_{n,\;m}=e^{-ik_y m} c_{n,\;m}(k_y) \label{Bloch2D_y}
\end{equation}
for $|k_y|\leq\pi/2$, where $c_{n,\;m+2}(k_y)=c_{n,\;m}(k_y)$ and
$c_{n,\;m}(k_y+\pi)=c_{n,\;m}(k_y)$. Therefore, the generalized
Harper equation becomes
\begin{eqnarray}\label{Harper_2D}
&&{\bf A}_n \vec{c}_n(k_y) + {\bf B}_n \vec{c}_{n+1}(k_y) + {\bf
B}_{n-1} \vec{c}_{n-1}(k_y) \nonumber \\
&& = E \vec{c}_n(k_y)
\end{eqnarray}
where $\vec{c}_n(k_y)=(c_{n,\;1}(k_y), c_{n,\;2}(k_y))^T$ and
\begin{equation}
 {\bf A}_n=\left( \begin{array}{cc}
  -(-1)^n \Delta_V & -t_1 \cos (\chi_n) e^{i\delta_n} \\
   -t_1 \cos (\chi_n) e^{-i\delta_n} & (-1)^n \Delta_V
  \end{array}\right)   \label{matrix_A}
\end{equation}
\begin{equation}
 {\bf B}_n=\left( \begin{array}{cc}
  -t_1 /2  & -t_2 \cos (\eta_n) \\
   -t_2 \cos (\eta_n) &  -t_1 /2
  \end{array}\right)   \label{matrix_B}
\end{equation}
with $\chi_n = n\phi + k_{y}$, $\delta_n = (-1)^n \pi\Delta_\phi$,
and $\eta_n = (n+1/2)\phi + k_{y}$. It can be easily checked that
\begin{equation}
 {\bf A}_{n+Q}={\bf A}_{n},~ {\bf B}_{n+Q}={\bf B}_{n},
\end{equation}
where $Q=q$ ($2q$) for an even (odd) integer $q$. Thus, the $n$ in
Eq.~(\ref{Harper_2D}) satisfies the condition $1\le n\le Q$.
Besides, because of the magnetic translation symmetry, the
primitive unit cell is consisted of $q$ ($Q$) plaquettes without
(with) checkerboard modulation.

The Bloch condition along the $x$ direction can be written as
\begin{equation}
c_{n,\;m}(k_y)=e^{-ik_{x} n}\psi_{n,\;m}(k_x, k_y)
\label{Bloch2D_x}
\end{equation}
for $|k_{x}|\leq\pi/Q$, where $\psi_{n+Q,\;m}(k_x,
k_y)=\psi_{n,\;m}(k_x, k_y)$ and $\psi_{n,\;m}(k_x+2\pi/Q,
k_y+\pi)=\psi_{n,\;m}(k_x, k_y)$. Now we only need to solve the
problem within the first magnetic Brillouin zone given by
$|k_{x}|\leq\pi/Q$ and $|k_{y}|\leq\pi/2$.

Thus we obtain the eigenvalue equation, ${\bf M}\Psi=E\Psi$, where
$\Psi=(\psi_{1,\;1},\psi_{1,\;2},\psi_{2,\;1},\psi_{2,\;2},\cdots,
\psi_{Q,\;1},\psi_{Q,\;2})^T$
and
\begin{widetext}
\begin{equation}
 {\bf M}=\left( \begin{array}{ccccccc}
  {\bf A}_{1} & {\bf B}_{1}e^{-ik_x} & 0 & \cdots & 0 & 0 &{\bf B}_{Q} e^{ik_x} \\
  {\bf B}_{1}e^{ik_x} & {\bf A}_{2} & {\bf B}_{2}e^{-ik_x} & \cdots & 0 & 0 & 0 \\
  0 & {\bf B}_{2}e^{ik_x} & {\bf A}_{3} & \cdots & 0 & 0 & 0 \\
  \cdots & \cdots & \cdots & \cdots & \cdots & \cdots & \cdots \\
  0 & 0 & 0 & \cdots & {\bf B}_{Q-2}e^{ik_x} &  {\bf A}_{Q-1} & {\bf B}_{Q-1}e^{-ik_x} \\
  {\bf B}_{Q}e^{-ik_x} & 0 & 0 & \cdots & 0  & {\bf B}_{Q-1}e^{ik_x} & {\bf A}_{Q}
  \end{array}\right).                            \label{matrix2D}
\end{equation}
\end{widetext}
We calculate the energy eigenvalues for all the values of
$\vec{k}$ in the first magnetic Brillouin zone, $|k_{x}|\leq\pi/Q$
and $|k_{y}|\leq\pi/2$, by directly diagonalizing the $2Q \times
2Q$ Hamiltonian matrix ${\bf M}(\vec{k})$. As indicated in Fig.~1,
the system has the checkerborad translational symmetry. That is,
the system is invariant under the lattice translation by two
lattice constants along either the $x$ or the $y$ directions, or
under the translation $(n,\;m)\rightarrow(n+1,\;m+1)$ along the
diagonal. Thus one expects that, under the above transformations,
the energy spectrum obtained by the eigenvalue problem with the
Hamiltonian matrix ${\bf M}(\vec{k})$ should remain the same,
which is not obvious as seen from Eq.~(\ref{matrix2D}). In
Appendix A, we prove that the Hamiltonian matrices before and
after the translations are identical up to a shift in $k_y$ and
thus give the same energy spectra.

We show that there are several general symmetries of the spectra
in the $E-\phi$ diagram, which can be used to reduce the amount of
calculations. Similar discussion for the systems without field
modulations can be found in Ref. [\onlinecite{hatsugai90}]. In the
following, the collecton of energy subbands at a flux $\phi$ (per
plaquette) modulated by $(\Delta_\phi,\Delta_V)$ is denoted by
$E(\phi,\Delta_\phi,\Delta_V)$. It has the following
symmetries:\cite{note1}

\begin{enumerate}
  \item \label{s1} $E(\phi,\Delta_\phi,\Delta_V)=E(-\phi,-\Delta_\phi,
  \Delta_V)$
  \\This follows from using two (three-dimensional) coordinate systems which are mirror
  images of each other with respect to the $x-y$ plane. The
  physics, and hence the energy spectra, should be the same
  in these two frames with opposite handnesses.

  \item \label{s2} $E(\phi,\Delta_\phi,\Delta_V)=E(-\phi,
  \Delta_\phi,\Delta_V)=E(-\phi,-\Delta_\phi,-\Delta_V)$
  \\The first equality follows from rotating the (three-dimensional)
  coordinate frame by 180 degrees around either the $x$-axis or the $y$-axis;
  the second is from shifting the coordinate by one
  plaquette along either the $x$-axis or the $y$-axis.

  \item \label{s3} $E(\phi,\Delta_\phi,\Delta_V)=E(\phi+2\phi_0,
  \Delta_\phi,\Delta_V)$
 \\This results from the following two facts: (i) the smallest
hopping loop for electrons encloses half of a plaquette; (ii) the
Aharonov-Bohm phase for this closed loop is unchanged after adding
one flux quantum to within this loop.

  \item \label{s4} $E(\phi,\Delta_\phi,\Delta_V)=\Re E(\phi+\phi_0,
  -\Delta_\phi,\Delta_V)$, where the operator $\Re$ flips the spectrum
  with respect to the horizontal $E=0$ line.
  \\This follows from the two transformations:
  (i) $f_{n,m} \rightarrow (-1)^{n+m}f_{n,m}$ and $\phi \rightarrow \phi+\phi_0$;
  (ii) $f_{n,m} \rightarrow f_{n,m+1}$ and $\Delta_\phi \rightarrow
  -\Delta_\phi$. It can be shown that the overall sign of the
  Hamiltonian in Eq.~(\ref{tight-binding}) changes after this two
  transformations, thus the spectra should have symmetry IV after
  shifting $\phi$ by one $\phi_0$ and reversing the direction
  of $\Delta_\phi$.

  \item \label{s5} $E(\phi,\Delta_\phi,\Delta_V)=E(\phi-\phi_0,\Delta_\phi-1,
  \Delta_V)$
  \\This is because of the freedom in shifting the fluxes,
  $(\phi_A,\phi_B) \rightarrow (\phi_A-2\phi_0,\phi_B)$,
  where $\phi_A=\phi+2\pi\Delta_\phi$ and $\phi_B=\phi-2\pi\Delta_\phi$
  are the fluxes through each plaquette of
  the $A$ and $B$ sublattices respectively.

\end{enumerate}

By combining symmetries I and II, it is not difficult to see that
$E(\phi,\Delta_\phi,\Delta_V)$ should remain unchanged when the
sign of {\it any} of its arguments, $\phi$, $\Delta_\phi$, or
$\Delta_V$, is changed. From symmetries III and II, we have
$E(\phi_0+\phi,\Delta_\phi,\Delta_V)=E(\phi-\phi_0,\Delta_\phi,\Delta_V)
=E(\phi_0-\phi,\Delta_\phi,\Delta_V)$.
That is, the distribution of $E(\phi)$ in the $E-\phi$ diagram has
a mirror symmetry with respect to the vertical line $\phi=\phi_0$.

Finally, we show that, in the $E-\phi$ diagram, it is sufficient
to plot the spectra within the range $0\leq\phi<\phi_0/2$ only.
The reason is as follows: from II, III, IV, and the freedom to
flip the signs of $\Delta_\phi$ and $\Delta_V$ without changing
the spectrum, it is clear that, for fixed values of $\Delta_\phi$
and $\Delta_V$, it suffices knowing the spectrum within the
interval $[0,\phi_0)$. Moreover, from IV and I, we have
$E(\phi_0/2+\phi,\Delta_\phi,\Delta_V)=\Re
E(\phi-\phi_0/2,-\Delta_\phi,\Delta_V)=\Re
E(\phi_0/2-\phi,\Delta_\phi,\Delta_V)$. Therefore, the spectrum
along the whole flux-coordinate can be determined by the $E(\phi)$
within the interval $[0,\phi_0/2)$.

\section{main features of the spectra}

In the discussion below, all energies are in units of $t_1$.
Besides $t_1$(=1), there are three adjustable parameters in the
present generalized Hofstadter model on a square lattice:
$\Delta_\phi$, $\Delta_V$, and $t_2$. It is impossible to show all
the results from the whole three-dimensional parameter
space.\cite{note2} Therefore, we selectively report on certain
sets of parameters with representative features. Notice that in
Refs.~[\onlinecite{shi97,oh99,ito99}], neither electric modulation
nor NNN hoppings has been considered and the parameter space is
one dimensional only.

First, we consider the effect of the checkerboard modulation on
the systems {\it without} NNN couplings $t_2$, which have several
symmetries {\it in addition to} the symmetries I$\sim$V listed
above.

\begin{enumerate}
  \item[\ref{s3}$^\prime$.] $E(\phi,\Delta_\phi,\Delta_V)=
  E(\phi+\phi_0,\Delta_\phi,\Delta_V)$
  \\When there is only NN hoppings, the period of the spectrum
  is one flux quantum since the smallest loop of hopping now is
  one plaquette, instead of half of the plaquette.

  \item[\ref{s4}$^\prime$.] $E(\phi,\Delta_\phi,\Delta_V)=
  \Re E(\phi,\Delta_\phi,\Delta_V)$
    \\This results from symmetries IV and III$^\prime$, followed by
    flipping the sign of $\Delta_\phi$, which would not change the
    spectrum. Thus, $E(\phi,\Delta_\phi,\Delta_V)$ is symmetric
    with respect to the horizontal $E=0$ line when $t_2$ =0.

  \item[\ref{s5}$^\prime$.] $E(\phi,\Delta_\phi,\Delta_V)=
  E(\phi-\phi_0/2,\Delta_\phi-1/2,\Delta_V)$
  \\The argument is similar to the one leading to V, except that
  now $\phi_A$ can be shifted by one $\phi_0$ without altering the
  Aharonov-Bohm phase of a closed-loop hopping.

\end{enumerate}

In Fig.~2, the spectrum for a checkerboard modulation with
$(\Delta_\phi,\Delta_V)=(0.1,0)$ is presented. The spectrum is
indeed symmetric with respect to the $E=0$ line, according to
symmetry IV$^{\prime}$. Furthermore, because of the symmetry
V$^{\prime}$ and the freedom to flip the signs of the arguments,
we have $E(\phi_0/4+\phi,\Delta_\phi,\Delta_V)
=E(\phi-\phi_0/4,1/2-\Delta_\phi,\Delta_V)
=E(\phi_0/4-\phi,\Delta_\phi-1/2,\Delta_V)$. Therefore, {\it after
being reflected} by the vertical line at $\phi_0/4$, Fig.~2 with
$\Delta_\phi=0.1$ is identical to the Hofstadter spectrum for
$\Delta_\phi=0.4$.\cite{note3}

In Fig.~3, a checkerboard modulation with
$(\Delta_\phi,\Delta_V)=(0.1,0.1)$ is considered. Without NNN
hoppings, this Hofstadter spectrum retains the same symmetries as
in Fig.~2. However, a distinctive $\phi$-independent energy gap
with a magnitude $E_g=2\Delta_V$ appears in the middle [also see
Fig.~6(a)]. This is true {\it with or without} adding the
modulation $\Delta_\phi$. First, it is not difficult to understand
why the spectrum splits to two groups in energy: they originate
from the two Bloch bands at $\phi=0$ due to the checkerboard
modulation of the scalar potential. What is surprising is that the
magnitude of the gap remains a constant for different $\phi$'s and
$\Delta_\phi$'s. It is no longer a constant as long as NNN
couplings are included (see Fig.~4). A proof of the existence of
the flux-independent gap is given in Appendix B, where it is shown
that there exists a very simple relation between the spectra with
and without electrostatic modulation $\Delta_V$. That is,
$E(\phi,\Delta_\phi,\Delta_V)=\pm[E(\phi,\Delta_\phi,0)^2+\Delta_V^2]^{1/2}$.
It can be checked that the spectra in Fig. 2 and Fig. 3 do obey
this relation in details.

The constancy of the energy gap {\it in the limit of small flux}
$\phi$ can be understood in the following semiclassical
picture.\cite{chang96} The energy bands with vanishing widths as
$\phi\rightarrow 0$ in Fig.~3 are the cyclotron energy levels of
the two parent bands at $\phi=0$, which have the energy
dispersions $E_\pm({\vec k})=\pm[(\cos k_x+\cos
k_y)^2+\Delta_V^2]^{1/2}$ {\it if} $\Delta_\phi=0$. It can be
shown that, near the two inner band edges with energies
$E_\pm=\pm\Delta_V$, the cyclotron effective masses approach
infinity. Therefore, the position of the lowest Landau level
approaches the lowest possible energy at the band edge and does
not depend on the uniform magnetic field.

When NNN couplings are included, the spectrum immediately lose the
mirror symmetry with respect to the horizontal $E=0$
line.\cite{hatsugai90} If only two of the three parameters are
nonzero, then the spectrum remains fractal but distorted. When all
three parameters, $\Delta_\phi,\Delta_V$, and $t_2$, are nonzero,
the subbands become significantly wider in most, but not all, of
the regions. A typical example is shown in Fig.~4. The extent of
widening varies as the parameters are varied. Because of the
widening, the electrons are more delocalized, and become more
mobile in transport.

There is a surprising exception to the asymmetry resulted from NNN
hoppings: the symmetry is restored again when $\Delta_\phi=0.5$,
even if {\it both} $\Delta_V$ and $t_2$ are nonzero. For example,
the symmetric spectrum shown in Fig.~5 is for ($
\Delta_\phi,\Delta_V,t_2)=(0.5,0.3,0.7)$. The existence of such a
symmetry can be proved as follows. From the symmetries III, IV,
and V listed above, and the freedom to flip the signs of the
arguments, it can be shown that $E(\phi,\Delta_\phi,\Delta_V)= \Re
E(\phi,1-\Delta_\phi,\Delta_V)$, which is a far less apparent
symmetry since it relates two systems with different strengths of
flux modulation. It is clear that when $\Delta_\phi=0.5$, the
spectrum has to be symmetric with respect to the line $E=0$.

In Fig.~6, we demonstrate how the continuous variation of
$\Delta_V$ and $\Delta_\phi$ influence the spectrum. The uniform
flux and the NNN couplings are fixed at the values of $p/q=2/5$
and $t_2=0$. In principle, there should be $Q=2q=10$ bands at this
value of the flux. However, in Fig.~6(a) with
$(\Delta_V,\Delta_\phi)=(x,0)$, where $x\in [0,1]$ is the value of
the $x$-coordinate, only 6 bands are observed. In fact, each of
the upper two and lower two bands is itself formed by two
overlapping subbands. We can also see that the band gap in the
middle is indeed proportional to $\Delta_V$, as mentioned earlier.
In Fig.~6(b), $(\Delta_V,\Delta_\phi)=(0,x)$, where $x\in [0,1]$
is again the value of the $x$-coordinate. There is almost no
similarity between (a) and (b). The overlapped subbands in
Fig.~6(a) are split by a nonzero $\Delta_\phi$ and become very
thin in most of the regions. On the other hand, the band in the
middle is thick and is actually composed of two subbands. The
increase of flux modulation also induces many band crossings. In
addition, there is an apparent symmetry
$E(\phi,\Delta_\phi,\Delta_V)=E(\phi,1-\Delta_\phi,\Delta_V)$. In
Fig.~6(c), both $\Delta_V$ and $\Delta_\phi$ are nonzero and have
the same numerical value. It has mixing features from (a) and (b),
but the magnitude of the energy gap in the middle is not altered
[comparing with (a)] by the nonzero $\Delta_\phi$. Such a
continuous tuning of the band structure might be realized in the
future using the optical lattices formed by quantum optical
means.\cite{jaksch03}

\section{summary}

The studies of Hofstadter spectrum have evolved from pure academic
curiosities to accessible experimental investigations. It is a
basic physics problem involving simple interplay between a lattice
and a magnetic field. Because of its general setting, it is not
surprising to find counterpart problems in different physical
systems, such as the quantum Hall system, the type-II
superconductivity, and the two-dimensional antiferromagnetism.
Motivated by a study on the frustrated antiferromagnetism, and the
recent experimental advances, we study the Hofstadter problem with
checkerboard modulations in details. In this paper, the spectra
are found to have several flux-related symmetries with respect to
the change of $\phi$ and $\Delta_\phi$. One unanticipated symmetry
occurs when $\Delta_\phi=1/2$. At that value, the spectrum are
symmetric with respect to the $E=0$ line even in the presence of
NNN hoppings. In the absence of NNN hoppings, we find a
flux-independent energy gap induced by electric modulations.
Furthermore, a simple connection between the spectra for {\it
bipartite} systems with and without electric modulation is
discovered. More detailed aspects of the spectra are not
investigated in this paper, however, such as the change of the
fractal measures in the $\Delta_\phi-\Delta_V-t_2$ parameter
space. Such a study would reveal different phases in this space,
as was done by Han and coworkers on the systems in a uniform
magnetic field.\cite{han94} The most general problem, when the
superlattices of modulation can have the symmetries of their own,
is considerally more involved. This study offers a starting point
for researches in this direction.

\acknowledgments M.C.C. and M.F.Y. acknowledge the financial
support from the National Science Council of Taiwan under Contract
Nos.  NSC 91-2112-M-003-019 and NSC 91-2112-M-029-007
respectively.

\appendix
\section{proof of the checkerboard-translation symmetry of the spectrum}
In this appendix, we show that the energy spectrum obtained from
Eq.~(\ref{matrix2D}) does respect the checkerboard translation
symmetry.

First, because there is no $m$-dependence of the matrix elements
in Eq.~(\ref{matrix2D}), the Hamiltonian matrix and therefore the
spectrum are unchanged under the lattice translation $m
\rightarrow m+2$ such that $\psi_{n,\;m}(k_x,
k_y)\rightarrow\psi_{n,\;m+2}(k_x, k_y)$. Second, from
Eqs.~(\ref{matrix_A}) and (\ref{matrix_B}), one can show that the
matrix elements in the Hamiltonian matrix satisfy the relations
${\bf A}_{n+2}(k_y)={\bf A}_{n}(k_y+2\phi)$ and ${\bf
B}_{n+2}(k_y)={\bf B}_{n}(k_y+2\phi)$. Therefore, under the
lattice translation $n \rightarrow n+2$ such that
$\psi_{n,\;m}(k_x, k_y)\rightarrow\psi_{n+2,\;m}(k_x, k_y)$, the
new Hamiltonian matrix for the eigenvalue problem after
transformation becomes identical to the original one with another
value of $k_y$, i.e., $k_y\rightarrow k_y+2\phi$. Thus the whole
energy spectrum within the first magnetic Brillouin zone remains
the same. Third, the matrix elements in the Hamiltonian matrix can
be shown to obey the following identities: $\sigma_x {\bf
A}_{n+1}(k_y) \sigma_x = {\bf A}_{n}(k_y+\phi)$ and $\sigma_x {\bf
B}_{n+1}(k_y) \sigma_x = {\bf B}_{n}(k_y+\phi)$, where $\sigma_x$
is the Pauli matrix. By using these identities, one can prove
that, under the lattice translation
$(n,\;m)\rightarrow(n+1,\;m+1)$ such that $\psi_{n,\;m}(k_x,
k_y)\rightarrow\psi_{n+1,\;m+1}(k_x, k_y)$, the new Hamiltonian
matrix again becomes identical to the original one with a shift
$k_y\rightarrow k_y+\phi$. Hence we conclude that the energy
spectrum is indeed invariant under the checkerborad translation.

\section{proof of the existence of the flux-independent energy gap}
For $t_2 =0$, our model is a nearest-neighbor-hopping model on a
bipartite lattice. Therefore, we can rewrite the Hamiltonian in
Eq.~(\ref{tight-binding}) as
\begin{equation}
H= (\{f^\dagger_A\},\{f^\dagger_B\}) \pmatrix{ \Delta_V I&{\cal
D}\cr {\cal D}^\dagger& -\Delta_V I\cr } \pmatrix{ \{f_A\}\cr
\{f_B\}\cr },
\end{equation}
where $I$ denotes the identity matrix, $\{f_A\}=\{f_{n,m}\ | \ n+m
{\rm \ is \ even} \}$ is a set of fermion operators for sublattice
$A$ and $\{f_B\}=\{f_{n,m}\ | \ n+m {\rm \ is \ odd}\}$ is for
sublattice $B$. When $\Delta_V=0$, the Schr$\rm{\ddot{o}}$dinger
equation is
\begin{eqnarray}
\pmatrix{ 0&{\cal D}\cr {\cal D}^\dagger&
0\cr}\pmatrix{\Phi_A\cr\Phi_B\cr}=E_0\pmatrix{\Phi_A\cr\Phi_B\cr},
\end{eqnarray}
where $E_0$ is the eigenvalue for the system with $\Delta_V=0$,
and $(\Phi_A,\Phi_B)^T$ is the corresponding eigenvector. From
them we can construct the eigenstates for the original problem:
\begin{eqnarray}
&&\Phi_+ \equiv \pmatrix{ \Delta_V + \sqrt{E_0^2 + \Delta_V^2}
\Phi_A \cr
{\cal D}^\dagger\Phi_A \cr }\\
&&\Phi_- \equiv \pmatrix{ {\cal D} \Phi_B \cr -\Delta_V -
\sqrt{E_0^2 + \Delta_V^2} \Phi_B \cr }
\end{eqnarray}
with the corresponding eigenvalues $E_\pm=\pm\sqrt{E_0^2 +
\Delta_V^2}$, because
\begin{equation}
\pmatrix{ \Delta_V I&{\cal D}\cr {\cal D}^\dagger& -\Delta_V I\cr
} \Phi_\pm = \pm \sqrt{E_0^2 + \Delta_V^2} \Phi_\pm .
\end{equation}
Therefore, the energy spectrum is symmetric with respect to the
horizontal $E=0$ line as mentioned in Sec.~III. The
positive-energy and the negative-energy parts are separated by an
energy gap $2\sqrt{|E_0|_{\rm min}^2 + \Delta_V^2}$, where
$|E_0|_{\rm min}$ is the minimum value of $|E_0|$ at given $\phi$
and $\Delta_\phi$. Since it has been shown that zero-energy modes
exist in the absence of $\Delta_V$ for all flux
values,\cite{hatsugai97} we have $|E_0|_{\rm min}=0$ for all
values of $\phi$ and $\Delta_\phi$. Consequently, the magnitude of
the energy gap in the presence of $\Delta_V$ should be
$2\Delta_V$, independent of the values of $\phi$ and
$\Delta_\phi$.

\newpage
\begin{figure}
\caption{ A square lattice with checkboard field modulation. The
magnetic flux through the left (right) plaquette is
$\phi+2\pi\Delta_\phi$ ($\phi-2\pi\Delta_\phi$). The scalar
potentials at the lattice points indicated by solid and empty dots
are $\Delta_V$ and $-\Delta_V$ respectively.
 }
\end{figure}
\begin{figure}
\caption{ The Hofstadter spectrum with the following parameters:
$(\Delta_\phi,\Delta_V,t_2)=(0.1,0,0)$.
 }
\end{figure}
\begin{figure}
\caption{ The Hofstadter spectrum with the following parameters:
$(\Delta_\phi,\Delta_V,t_2)=(0.1,0.1,0)$.
 }
\end{figure}
\begin{figure}
\caption{ The Hofstadter spectrum with the following parameters:
$(\Delta_\phi,\Delta_V,t_2)=(0.1,0.1,0.5)$.
 }
\end{figure}
\begin{figure}
\caption{ The Hofstadter spectrum with the following parameters:
$(\Delta_\phi,\Delta_V,t_2)=(0.5,0.3,0.7)$.
 }
\end{figure}
\begin{figure}
\caption{ The variation of band widths as (a) $\Delta_V$, (b)
$\Delta_\phi$, and (c) both $\Delta_V$ and $\Delta_\phi$ are
tuned. The values of $\phi/2\pi=2/5$ and $t_2=0$ are fixed.
 }
\end{figure}


\begin{thebibliography}{99}

\bibitem{hofstadter76}
 D.~R. Hofstadter, Phys. Rev. {\bf 14}, 2239 (1976).
\bibitem{thouless83}
 D.~J. Thouless and Q. Niu, J. Phys. A, Math. Gen. {\bf 16}, 1911
 (1983); D. Dominguez, C. Wiecko, and J. Jose, Phys. Rev. B {\bf
 45}, 13919 (1992).
\bibitem{claro79}
 F.~H. Claro and G.~H. Wannier, Phys. Rev. B {\bf 19}, 6068 (1979).
\bibitem{rammal85}
 R. Rammal, J. Phys. (Paris) {\bf 46}, 1345 (1985).
\bibitem{xiao02}
 Y. Xiao, V. Pelletier, P.~M. Chaikin, and D.~A. Huse, Phys. Rev. B
 {\bf 67}, 104505 (2003).
\bibitem{vidal98}
 J. Vidal, R. Mosseri, and B. Doucot, Phys. Rev. Lett. {\bf 81},
 5888 (1998).
\bibitem{hatsugai90}
 Y. Hatsugai and M. Kohomoto, Phys. Rev. B {\bf 42}, 4282 (1990).
\bibitem{lee94}
 M.~Y. Lee, M.~C. Chang, and T.~M. Hong, Phys. Rev. B {\bf 57},
 11895 (1998).
\bibitem{han94}
 J.~H. Han, D.~J. Thouless, H. Hiramoto, and M. Kohmoto, Phys. Rev.
 B {\bf 50}, 11365 (1994).
\bibitem{barelli92}
 A. Barelli and R. Fleckinger, Phys. Rev. B {\bf 46}, 11559
 (1992).
\bibitem{gerhardts96}
 P.~D. Ye {\it et al}, Appl. Phys. Lett. {\bf 67}, 1441 (1995);
 R.~R. Gerhardts, D. Pfannkuche, and V. Gudmundsson, Phys. Rev. B
 {\bf 53}, 9591 (1996).
\bibitem{oh95}
 G.~Y. Oh, J.~Jang, and M.~H. Lee, J. Korean Phys. Soc. {\bf 28},
 79 (1995); G.~Y. Oh and M.~H. Lee, Phys. Rev. B {\bf 53}, 1225
 (1996); P. Fekete and G. Gumbs, J. Phys.: Condens. Matter {\bf 11}, 5475
 (1999).
\bibitem{shi97}
 Q.~W. Shi and K.~Y. Szeto, Phys. Rev. B {\bf 56}, 9251 (1997).
\bibitem{oh99}
 G.~Y. Oh, Phys. Rev. B {\bf 60}, 1939 (1999).
\bibitem{ito99}
 S. Ito, M. Ando, S. Katsumoto, and Y. Iye, J. Phys. Soc. Japan
 {\bf 68}, 3158 (1999); M. Ando, S. Ito, S. Katsumoto, and Y. Iye,
 {\it ibid} {\bf 68}, 3462 (1999).
\bibitem{kuhl98}
 U. Kuhl and H.~J. Stockmann, Phys. Rev. Lett. {\bf 80}, 3232
 (1998).
\bibitem{richoux02}
 O. Richoux and V. Pagneux, Europhys. Lett. {\bf 59}, 34 (2002).
\bibitem{albrecht01}
 C. Albrecht, J.~H. Smet, K. von Klitzing, D. Weiss, V. Umansky,
 and H. Schweizer, Phys. Rev. Lett. {\bf 86}, 147 (2001).
\bibitem{morita01}
 Y. Morita and Y. Hatsugai, Phys. Rev. Lett. {\bf 86}, 151 (2001);
 H.~K. Nguyen and S. Chakravarty, Phys. Rev. B {\bf 65}, 180519 (2002).
\bibitem{chamon03}
 C. Chamon, M. Oshikawa, and I. Affleck, cond-mat/0305121.
\bibitem{jaksch03}
 D. Jaksch and P. Zoller, New J. Phys. {\bf 5}, 56 (2003).
\bibitem{misguich01}
 G. Misguich, Th. Jolicoeur, and S.~M. Girvin, Phys. Rev. Lett.
 {\bf 87}, 097203 (2001).
\bibitem{chang02}
 M.~C. Chang and M.~F. Yang, Phys. Rev. B {\bf 66}, 184416 (2002).
\bibitem{note0}
 The antiferromagnetic couplings $J_1$ and $J_2$ in the Heisenberg model
 are equal to $-t_1$ and $-t_2$ respectively. The non-zero $J_2$ coupling
 results in a {\it frustrated} antiferromagnetic system. After a mean-field
 approximation, it can be shown that $\Delta_V=4(J_1-J_2)\Delta_\phi$ (see Sec.
 II for the definitions of $\Delta_\phi$ and $\Delta_V$).
 Therefore, there is only one parameter from the modulation. In
 this paper, we treat $\Delta_\phi$ and $\Delta_V$ as independent
 quantities.
\bibitem{note1}
 Symmetries I and II remain valid even if there are couplings
 beyond next-nearest neighbors.
\bibitem{note2}
 We have carried out calculations with the following sets of
 parameters: $(\Delta_\phi,\Delta_V,t_2)=(0.1i,0,0.1j),(0,0.1i,0.1j)$,
 and $(0.1i,0.1i,0.1j)$, where $i$ and $j$ are non-negative integers
 below 5 and 9 respectively. That is, 180 Hofstadter spectra have
 been generated.
\bibitem{note3}
 The parameter $(2/\pi)\beta$ in Ref.~13 is the same as our $\Delta_\phi$.
 Therefore, its Fig.~1(b) with $\beta=0.2\pi$ is the figure with
 $\Delta_\phi=0.4$. However, the edges of subbands for a few simple fractions,
 e.g., $p/q=2/3$, in their Fig.~1(b) fail to align with nearby
 edges when the uniform flux is slightly varied.
\bibitem{chang96}
 M.~C. Chang and Q. Niu, Phys. Rev. B {\bf 53}, 7010 (1996).
\bibitem{hatsugai97}
Y. Hatsugai, X. G. Wen, and M. Kohmoto, Phys. Rev. B {\bf 56},
1061 (1997).

\end{thebibliography}
\end{document}